# Open-Source Synthesizable Analog Blocks for High-Speed Link Designs: 20-GS/s 5b ENOB Analog-to-Digital Converter and 5-GHz Phase Interpolator

Sung-Jin Kim, Zachary Myers, Steven Herbst, ByongChan Lim, Mark Horowitz
Stanford University, CA, U.S.A.
sjkim85@stanford.edu

**Abstract**

Using digital standard cells and digital place-and-route (PnR) tools, we created a 20 GS/s, 8-bit analog-to-digital converter (ADC) for use in high-speed serial link applications with an ENOB of 5.6, a DNL of 0.96 LSB, and an INL of 2.39 LSB, which dissipated 175 mW in 0.102 mm$^2$ in a 16nm technology. The design is entirely described by HDL so that it can be ported to other processes with minimal effort and shared as open source.

## Introduction

In recent years, many open-source digital circuit generators have been developed to make it easier to create ASICs for demanding applications such as machine learning. Unfortunately, high-speed transceivers, which traditionally require precise analog circuits, are harder to open-source. As a result, designers still have to obtain proprietary transceiver IPs, which are expensive, hard to verify, and inflexible. To break this bottleneck, this paper describes the design and measurement of our open-source, portable generator for a high-speed link transceiver [1].

We can open-source this design because it uses digital standard cells and digital PnR tools, and does not rely on precision analog design. In fact, it takes advantage of circuit variations in its novel stochastic time-to-digital converter (STDC) that, in conjunction with a synthesizable phase interpolator (PI), is used to build a time-interleaved ADC that can run to 20 GS/s. While the idea of taking advantage of variation in an ADC is not new [2,3,4], our approach is much more power efficient than previous work.

## The proposed ADC

Fig. 1 shows the architecture of the proposed ADC. It consists of two voltage-to-time (V2T) converters, a V2T clock generator, a phase folder (PF), and an STDC. Each V2T samples its input voltage to $C_S$ at $\Phi_1$. At $\Phi_2$, the sampled voltage ($V_C$) is discharged toward ground at a constant rate by a current source; a digital buffer generates an edge when the $V_C$ crosses its logic threshold voltage. The first gate of that buffer is a 3-input NOR to position the threshold below VDD/2. For the current source, a thick-gate cascode device is used to achieve sufficient output resistance and eliminate the need for a second bias voltage. The V2T clock generator provides a slightly early phase ($\Phi_{1e}$) for bottom-plate sampling and a slightly late phase ($\Phi_{2l}$) for stable settling. The V2Ts effectively encode the input signal as the time difference between their outputs $T_{IN}P$ and $T_{IN}N$, which the PF folds into an unsigned pulse ($P_{IN}$), along with a sign bit. $P_{IN}$ includes a small offset delay $D_{offset}$ so as to achieve a minimum width of about 100 ps.

The STDC quantizes the width of $P_{IN}$ by counting the number of delayed clock edges it contains. These clock edges are generated from 255 non-precise unit inverters, and an adder tree and unfolder generate the signed binary result, as shown in Fig. 2. A divided clock from the V2T clock generator propagates through the inverter chain, generating a distribution of clock edges that is quasi-uniform in time because the period of the clock is uncorrelated to the delay of each inverter [5]. Due to the inherent immunity of the STDC against device mismatch, jitter, and PVT variation, it is a good fit for an automated PnR flow. The unfolder removes offset and converts the unsigned value back to a signed value according to the sign decision from the PF. The absolute value of the offset that should be subtracted is calculated through a background adaptation loop based on a histogram of output codes from the ADC (ADC$_{OUT}$).

Fig. 3(a) shows the PI used to generate and adjust sampling clocks for the multi-channel ADC. The delay chain generates 32 delayed phases of input clock ($\varphi_n$). The mux network selects one out of 16 odd phases ($\varphi^B_{2n-1}$) for the first input of the phase blender (ph$_{SEL1}$) and one out of 16 even phases ($\varphi^B_{2n}$) for the second input of the phase blender (ph$_{SEL2}$). Additional circuitry is added before the muxes to make the operation robust to variations. Arbiters are used to find the number of delays in a clock cycle, and this information is used to control the phase mixers (1-bit phase blenders) connected to every node of the delay chain. The one mixer at the clock boundary blends its input edge with the input clock edge to generate an average edge between them, while the rest of the mixers act as buffers to achieve monotonic phase rotation across this edge as shown in Fig. 3(b)(c). The encoder controls the phase mixers, the mux network, and the phase blender based on the input control code and the quantized period from the arbiters. The phase blender is implemented by 16 output-shorted muxes. Unfortunately, the monotonicity of the PI may be jeopardized by automated PnR when the mismatch among accumulated path delays exceeds the unit delay of the delay chain ($T_D$). This situation is detected and corrected after fabrication by adding an arbiter to the input of the phase blender and adjusting the strength of the delay buffer for the corresponding path.

The overall architecture of the proposed multi-channel ADC is shown in Fig. 4. Sixteen ADCs are time-interleaved to support a 20 GS/s conversion rate. Four groups of ADCs, each of which consists of four ADC slices, are connected to individual switches (SW$_0$-SW$_3$) without inter-stage buffers, forming a two-stage passive track and hold (T&H) structure (front-end switches in each ADC slice act as the second-stage). Four PIs take a quarter-rate input clock and generate 5 GHz quadrature sampling clock phases for the first T&H switches. Each PI has an independent control dedicated for canceling clock skew among sampling phases. Quantized data from each ADC slice are synchronized by a double flop aligner to become the final output. Look up table based static non-linearity calibration is done supported by an on-chip SRAM. An analog voltage generator based on logic gates provides bias voltages for the current source cells in all V2Ts.

It is worthwhile to note that the design is described entirely in SystemVerilog and was verified through Verilog simulation. For the few building blocks that could not be directly simulated by the Verilog simulator, such as the phase blender, V2T, and bias generator, event-driven functional models were generated through DaVE, an automated model generation flow from Stanford [6]. This enables over 100k times faster simulation time with sufficiently accurate results as compared to conventional SPICE simulation. The physical design of the proposed multi-channel ADC was done with a standard digital PnR flow with no manual layout, aside from two custom cells (a switch cell and the current source cell in the V2T).

## Measurement results

A prototype chip was fabricated in the TSMC 16nm FinFET technology. Fig. 5 illustrates measurement results of the proposed

PI. Its transfer function, measured by an on-chip delay monitoring circuit based on uncorrelated sampling [7], is monotonic with a resolution of 0.7 ps. The measured performance of the ADC is shown in Fig. 6, demonstrating a DNL of 0.95 LSB and an INL of 2.39 LSB. The ENOB is 5.6 at low frequencies, falling to 2.7 at Nyquist. The ENOB frequency dependence is mainly due to input power loss induced by package network (676pin FCBGA) which is not optimized for high frequency testing. The areas of an ADC slice and PI are 80μm x 40μm and 25μm x 80μm and they consume 8.6mW and 9.6mW respectively. The area of the 16-channel ADC including 4 PIs is 300μm x 340μm and the total power consumption is 175mW under a 0.9V supply. The performance of our ADC is summarized and compared with published papers in Fig. 7.

## Conclusions

In this paper, we presented work towards a portable analog generator for a high-speed link transceiver: namely, open-source, synthesizable designs for a 20-GS/s 8-bit ADC and a four-phase 5 GHz PI. By using circuit topologies that are insensitive to exact device parameters and whose analog constraints can be expressed in the time domain, we were able to make extensive use of digital tools for synthesis, layout, and simulation. Finally, we validated these techniques by fabricating a prototype chip in a 16nm technology.

## Acknowledgements
This work is supported by Stanford SystemX, DARPA POSH, a Hertz Fellowship, and a Stanford Graduate Fellowship.

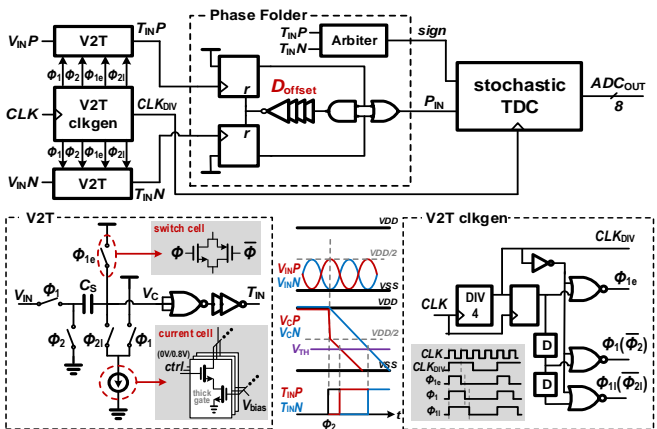

Fig. 1: Proposed fully synthesizable ADC slice based on a stochastic TDC.

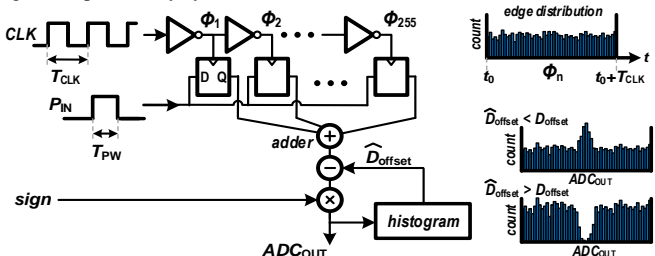

Fig. 2: Stochastic time-to-digital converter and unfolder.

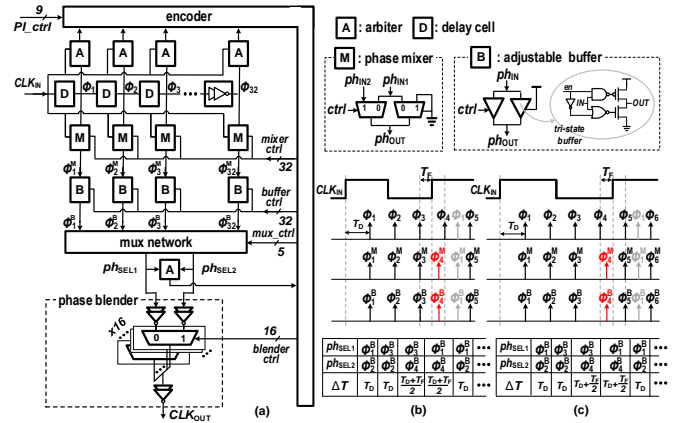

Fig. 3: Proposed fully synthesizable PI slice based on a digital phase blender.

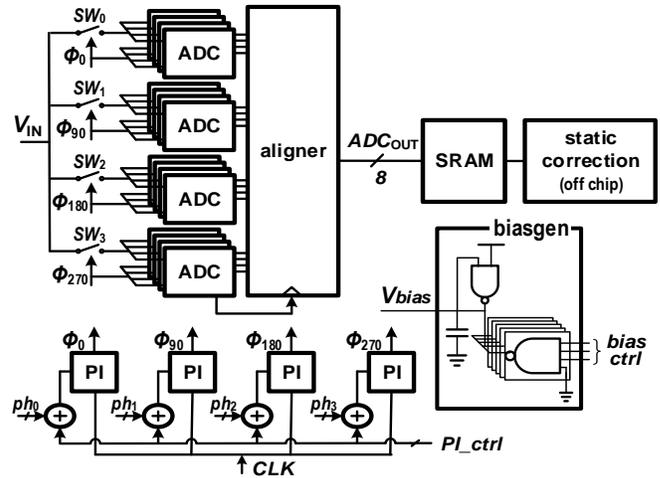

Fig. 4: Proposed 16-channel time-interleaved ADC architecture.

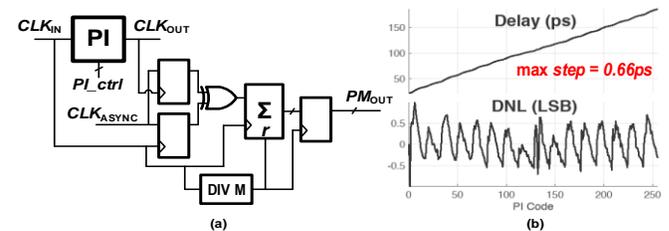

Fig. 5: Measured results of the PI.

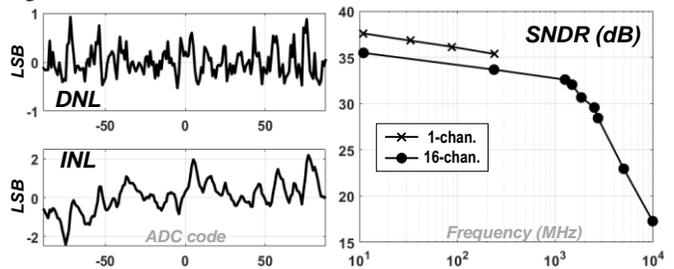

Fig. 6: Measured results of the ADC.

|  | [2] | [3] | [4] | This Work |
|---|---|---|---|---|
| Type | Stochastic Flash ADC | Stochastic Flash ADC | Stochastic Flash ADC | Stochastic Flash TDC |
| Process (nm) | 90 | 90 | 130 | 16 |
| Supply (V) | 1.2 | 1.2 | 1.0 | 0.9 |
| Normalized Input Range (VDD) | 0.04 | 0.12 | 0.4 | 0.5 |
| Speed (GHz) | 1.5 | 0.21 | 0.32 | 1.25 / 20 |
| ENOB | 3.7 | 5.69 | 5.17 | 5.9 / 5.6 |
| Power (mW) | 23 | 34.8 | 87 | 8.6 / 175 |
| FoM (pJ/b) | 1.18 | 3.2 | 7.55 | 0.12 / 0.18 |
| Area ($mm^2$) | 0.04 | 0.18 | 0.51 | 0.003 / 0.1 |
| Synthesizable | No | Yes | Yes | Yes |

Fig. 7: Test chip layout and performance comparison.